\documentclass[conference]{IEEEtran}

\usepackage[T1]{fontenc}
\usepackage{amsmath}
\usepackage{amssymb}
\usepackage{stmaryrd}
\usepackage{epic,eepic}
\usepackage{theorem}
\usepackage{pifont}  
\usepackage{euscript}
\usepackage{calc}
\usepackage[titles]{tocloft}

\usepackage[usenames]{color}          
\usepackage{xcolor}
\definecolor{Brown}{rgb}{0.55,0.0,0.10}
\definecolor{dgreen}{rgb}{0.00,0.56,0.00}
\definecolor{vertmoinsfonce}{rgb}{0.00,0.50,0.00}
\definecolor{vert}{rgb}{0.00,0.60,0.00}
\definecolor{llightggray}{rgb}{0.97,0.97,0.97}
\definecolor{lightggray}{rgb}{0.9,0.9,0.9}
\definecolor{ggray}{rgb}{0.5,0.5,0.5}
\definecolor{darkggray}{rgb}{0.25,0.25,0.25}
\definecolor{ddarkggray}{rgb}{0.1,0.1,0.1}
\definecolor{bleu}{rgb}{0.00,0.00,1.00}

\usepackage{epsf}           
\usepackage{graphicx}       
\usepackage{epsfig}         
\usepackage{pstricks}
\usepackage{pstricks-add}
\usepackage{pst-plot}
\usepackage{dsfont}
\theoremheaderfont{\color{black}\normalfont\bfseries}

\newtheorem{theorem}{Theorem}[section]
\newtheorem{definition}[theorem]{Definition}
\newtheorem{corollary}[theorem]{Corollary}
\newtheorem{proposition}[theorem]{Proposition}
\newtheorem{example}[theorem]{Example}

\theoremstyle{plain}{\theorembodyfont{\rmfamily}%
}
\theoremstyle{plain}{\theorembodyfont{\rmfamily}%
}
\theoremstyle{plain}{
\theorembodyfont{\rmfamily}

	}

\newcommand{\R}{\mathbb{R}}

\newcommand{\N}{\mathbb{N}}
\newcommand{\PPP}{\mathbb{P}}
\newcommand{\E}{\mathbb{E}}
\newcommand{\Q}{\mathbb{Q}}

\font\dsrom=dsrom10 scaled 1200 \def \indic{\textrm{\dsrom{1}}}
\newcommand{\UN}{\indic}

\newcommand{\C}{\mathcal{C}}

\newcommand{\QQ}{\mathcal{Q}}

\newcommand{\D}{\mathcal{D}}

\newcommand{\PP}{\mathcal{P}}

\newcommand{\mc}{\mathcal}

{\Large\normalsize}
{\Large\normalsize}

\ifCLASSINFOpdf
\else
\fi
\begin{document}

\title{Correlation between Channel State and Information Source with Empirical Coordination Constraint}

%
%
%

\author{\IEEEauthorblockN{Ma\"{e}l Le Treust}\\
\IEEEauthorblockA{
ETIS, CNRS UMR8051, ENSEA, Université de Cergy-Pontoise,\\
6, avenue du Ponceau,\\
95014 CERGY-PONTOISE CEDEX,\\
FRANCE\\
Email: mael.le-treust@ensea.fr}
}



\maketitle

\begin{abstract}
Correlation between channel state and source symbol is under investigation for a joint source-channel coding problem. 
We investigate simultaneously the lossless transmission of information and the empirical coordination of channel inputs with the symbols of source and states.
Empirical coordination is achievable if the sequences of source symbols, channel states, channel inputs and channel outputs are jointly typical for a target joint probability distribution. We characterize the joint distributions that are achievable under lossless decoding constraint. The performance of the coordination is evaluated by an objective function. For example, we determine the minimal distortion between symbols of source and channel inputs for lossless decoding.
We show that the correlation source/channel state improves the feasibility of the transmission.


\end{abstract}




%
\IEEEpeerreviewmaketitle

\section{Introduction}\label{sec:Introduction}

State-dependent channels with state information at the encoder have been widely studied since the publication of Gel'fand Pinsker's article \cite{gelfand-it-1980}. The authors characterize  the channel capacity in the general case and Costa evaluates it for channels with additive white Gaussian noise in \cite{costa-it-1983}. Interestingly, additional noise does not lower the channel capacity while it is observed by the encoder. In this setting, the encoder has non-causal state information, \textit{i.e.} it observes the whole sequence of channel states. This hypothesis is relevant for problems such as computer memory with defect  \cite{heegard-it-1983} in which the encoder observes the sequence of defects or digital watermarking \cite{MoulinOSullivan03} since the sequence of states is a watermarked image/data. Non-causal state information is also appropriate for the problem of empirical coordination \cite{LeTreust(EmpiricalCoordination)14}, since the sequence of source symbol acts as channel states. More recently, the notion of "state amplification" was introduced in \cite{KimSituvongCover08}. In this framework, the decoder is also interested in acquiring some information about the channel state. The amount of such information is measured by the  "uncertainty reduction rate". The authors characterize the optimal trade-off region between reliable data rate and uncertainty reduction rate. As a special case,  they consider that the encoder only transmit the sequence of channel-states.

In this paper, we investigate the problem of correlation between channel state and source symbol with a constraint of empirical coordination. Although the problem of correlation state/source is different from the problem of state amplification, both problems coincide when considering the transmission of channel states only. The reliability criteria we consider is based on empirical coordination \cite{WeissmanOrdentlich05}, \cite{KramerSavari07}, \cite{CuffPermuterCover10}, also referred as empirical correlation. An error occurs if sequences of source symbols, channel states, channel inputs and channel outputs are not jointly typical for a target joint probability, or if the decoding is not lossless. 
This result applies to coordination games \cite{LeTreust(EmpiricalCoordination)14}, in which the encoder aims at maximizing an objective function that depends on symbols of source and channel while transmitting lossless information to the decoder. 

\begin{figure}[!ht]
\begin{center}
\psset{xunit=0.9cm,yunit=0.9cm}
\begin{pspicture}(0,-0.65)(8.5,1.15)
\pscircle(0,0.5){0.45}
\psframe(2,0)(3,1)
\pscircle(5,0.5){0.45}
\psframe(7,0)(8,1)
\psline[linewidth=1pt]{->}(0.5,0.5)(2,0.5)
\psline[linewidth=1pt]{->}(3,0.5)(4.5,0.5)
\psline[linewidth=1pt]{->}(5.5,0.5)(7,0.5)
\psline[linewidth=1pt]{->}(8,0.5)(9,0.5)
\psline[linewidth=1pt]{->}(0,0)(0,-0.5)(5,-0.5)(5,0)
\psline[linewidth=1pt]{->}(2.5,-0.5)(2.5,0)
\rput[u](1,0.8){$U^n$}
\rput[u](1,-0.2){$S^n$}
\rput[u](3.75,0.8){$X^n$}
\rput[u](6.25,0.8){$Y^n$}
\rput[u](8.5,0.8){$\hat{U}^n$}
\rput(0,0.5){$\PP_{\sf{us}}$}
\rput(2.5,0.5){$\C$}
\rput(5,0.5){$\mc{T}$}
\rput(7.5,0.5){$\D$}
\end{pspicture}
\caption{Channel state $S$  is correlated with information source $U$. Encoder $\C$ and decoder $\D$ implement a coding scheme such that sequences of symbols $(U^n ,S^n,X^n,Y^n,\hat{U}^n) \in A_{\varepsilon}^{{\star}{n}}(\QQ)$ are jointly typical for the target probability distribution $\QQ(u,s,x,y,\hat{u})$ and the decoding is lossless. }
\label{fig:CorrelGPsideInfoEnc}
\end{center}
\end{figure}
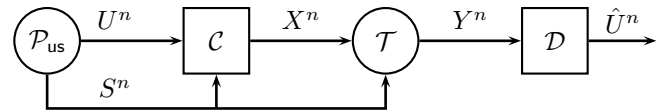

Channel model and definitions are presented in Sec. \ref{sec:ModelDefinition}. In Sec. \ref{sec:CorrelGPsideInfoEnc}, we characterize the set of achievable empirical distribution. In Sec. \ref{sec:PreviousWork}, we compare this result to previous works and to the case of non-correlated channel state/information source. Sec.\ref{sec:CorrelExampleBSAC} provides numerical results that illustrate the benefit of taking into account the correlation between channel state and information source. The conclusion is stated in Sec. \ref{sec:Conclusion} and  sketches of proof of the main result are presented in App. \ref{sec:ProofAchievability} and \ref{sec:ProofConverse}.

%

\section{System Model}\label{sec:ModelDefinition}

The problem under investigation is depicted in Fig. \ref{fig:CorrelGPsideInfoEnc}. Capital letter $U$ denotes the random variable and lowercase letter $u\in\mc{U}$ denotes the realization. We denote by $U^n$, $S^n$, $X^n$, $Y^n$,  $\hat{U}^n$, the sequences of random variables of the source symbols $u^n=(u_1,\ldots,u_n)\in\mc{U}^n$,  of channel state $s^n\in\mc{S}^n$, of channel inputs $x^n\in\mc{X}^n$, of channel outputs $y^n\in\mc{Y}^n$ and of outputs of the decoder $\hat{u}^n\in\mc{U}^n$. We assume the sets $\mc{U}$, $\mc{S}$, $\mc{X}$, $\mc{Y}$ are discrete and $\mc{U}^n$ denotes the $n$-time cartesian product of set $\mc{U}$.  Notation $\Delta(\mc{X})$ stands for the set of  probability distributions $\PP(X)$ over $\mc{X}$. The variational distance between probabilities $\QQ$ and $\PP$ is denoted by $||\QQ - \PP||_{\sf{v}}= 1/2\cdot \sum_{x\in\mc{X}} |\QQ(x) - \PP(x)|$.
Notation $\UN(\hat{u}|u)$ 
denotes the indicator function that equals 1 if $\hat{u} = u$ and 0 otherwise.  The Markov chain property is denoted by $Y  -\!\!\!\!\minuso\!\!\!\!-X    -\!\!\!\!\minuso\!\!\!\!-  U$ and holds if for all $(u,x,y)$ we have $\PPP(y|x,u) = \PPP(y|x)$. The information source and channel states are i.i.d. distributed with $\PP_{\sf{us}}$ and the channel is memoryless  with transition probability $\mc{T}_{\sf{y|xs}}$. The statistics of $\PP_{\sf{us}}$ and $\mc{T}_{\sf{y|xs}}$ are known by both encoder $\C$ and decoder $\D$.

Coding Process: A sequence of source symbols $u^n\in\mc{U}^n$ and channel states $s^n\in\mc{S}^n$ are drawn from the i.i.d. probability distribution defined by equation \eqref{eq:SourceProba}. Encoder $\C$ observes $u^n\in\mc{U}^n$, $s^n\in\mc{S}^n$  non-causally and sends a sequence of channel inputs $x^n\in\mc{X}^n$. The sequence of channel outputs $y^n \in \mc{Y}^n$ is drawn according to the discrete and memoryless transition probability defined by equation (\ref{eq:TransitionProba}).
\begin{eqnarray}
\mc{P}_{\sf{us}}^{\otimes n}(u^n,s^n)= \prod_{i=1}^n \mc{P}_{\sf{us}}(u_i,s_i),\label{eq:SourceProba}\\
\mc{T}^{\otimes n}(y^n|x^n,s^n)= \prod_{i=1}^n \mc{T}(y_i|x_i,s_i).\label{eq:TransitionProba}
\end{eqnarray}
The decoder observes the sequence of channel outputs $y^n\in\mc{Y}^n$ and returns a sequence $\hat{u}^n\in\mc{U}^n$. We consider lossless decoding: the sequence of output of the decoder $\hat{u}^n\in\mc{U}^n$ should be equal to the source sequence $\hat{u}^n = u^n \in\mc{U}^n$. 

The objective of this work is to characterize the set of empirical distributions $\QQ \in \Delta(\mc{U} \times \mc{S} \times \mc{X}  \times \mc{Y}  \times \mc{U}  ) $ that are achievable, \textit{i.e.} for which the decoding is lossless and the encoder and decoder can implement sequences of symbols  $(U^n ,S^n,X^n,Y^n,\hat{U}^n) \in A_{\varepsilon}^{{\star}{n}}(\QQ)$ that are jointly typical for probability distribution $\QQ$ with high probability. The definition of typical sequence is stated pp. 27 in \cite{ElGammalKim(book)11}.  In this setting, the sequence of channel inputs $X^n$ must be coordinated (\textit{i.e.} jointly typical) with $(U^n ,S^n)$. The performance of the coordination is evaluated by an objective function  $\nu : \mc{U} \times \mc{S} \times \mc{X}  \times \mc{Y}  \mapsto \R$, as stated in \cite{LeTreust(EmpiricalCoordination)14}. For example, we characterize the minimal distortion level $d(u,x)$ between the source symbol $u\in \mc{U}$ and channel input $x\in\mc{X}$ under lossless decoding constraint. Error probability evaluates simultaneously the lossless transmission of information and the empirical coordination of symbols of source, state, channel input, channel output and decoder output. Since the decoding is lossless, a joint distribution $\QQ \in \Delta(\mc{U} \times \mc{S} \times \mc{X}  \times \mc{Y}  \times \mc{U}  ) $ is achievable if the marginal probability distribution $\QQ(\hat{u}|u)= \UN(\hat{u}|u)$ is equal to the indicator function. However, the lossless decoding condition is more restrictive  than marginal condition $\QQ(\hat{u}|u)= \UN(\hat{u}|u)$, since sequences $\hat{U}^n = U^n$ should be strictly equal.

\begin{definition}\label{def:Code}
A  code $c\in\mc{C}(n)$ with non-causal encoder is a pair of functions $c=(f,g)$ defined by equations \eqref{eq:1CausalCodeSource1}-\eqref{eq:1CausalCodeSource2}.
\begin{eqnarray}
&f:&  \mc{U}^n \times \mc{S}^n  \longrightarrow \mc{X}^n ,\label{eq:1CausalCodeSource1}\\
&g:&  \mc{Y}^n  \longrightarrow \mc{U}^n.\label{eq:1CausalCodeSource2}
\end{eqnarray}
The empirical distribution ${Q}^n \in \Delta(\mc{U}\times \mc{S} \times \mc{X}\times \mc{Y} \times\mc{U})$ of sequences $(u^n,s^n,x^n,y^n,\hat{u}^n)\in \mc{U}^n\times \mc{S}^n\times \mc{X}^n \times \mc{Y}^n \times\mc{U}^n$ is defined by equation \eqref{eq:EmpiricalDistribution} where $\textsf{N}(u|u^n)$ denotes the number of occurrence of symbol $u \in \mc{U}$ in sequence $u^n \in \mc{U}^n$.
\begin{eqnarray}
{Q}^n(u,s,x,y,\hat{u}) &=& \frac{\textsf{N}(u,s,x,y,\hat{u}|u^n,s^n,x^n,y^n,\hat{u}^n)}{n},  \nonumber \\
 \forall  (u,s,x,y,\hat{u})& \in& \mc{U}\times \mc{S}\times \mc{X} \times \mc{Y} \times\mc{U}. \label{eq:EmpiricalDistribution}
\end{eqnarray}
Fix a target probability distribution $\QQ \in \Delta(\mc{U} \times \mc{S}\times \mc{X} \times \mc{Y} \times \mc{U}  )$, the error probability of the code $c\in\mc{C}(n)$ is defined by equation \eqref{eq:ErrorProba}.
\begin{eqnarray}
\PP_{\textsf{e}}(c) = \PP_c\bigg(\Big|\Big|Q^n - \QQ \Big|\Big|_{\sf{v}}\geq \varepsilon\bigg)+ \PP_c\bigg( U^n \neq \hat{U}^n \bigg),\label{eq:ErrorProba}
\end{eqnarray}
where $Q^n \in \Delta(\mc{U}\times \mc{S}\times \mc{X}\times \mc{Y} \times\mc{U})$ is the random variable of the empirical distribution of the sequences of symbols $(U^n,S^n,X^n,Y^n,\hat{U}^n)\in \mc{U}^n\times \mc{S}^n \times  \mc{X}^n \times \mc{Y}^n \times\mc{U}^n$ induced by the code $c \in \mc{C}(n)$ and probability distributions $\PP_{\sf{us}}$, $\mc{T}_{\sf{y|xs}}$.
\end{definition}

\begin{definition}
A probability distribution $\QQ \in  \Delta(\mc{U} \times \mc{S}\times \mc{X} \times \mc{Y} \times \mc{U}  )$ is achievable if for all $\varepsilon>0$, there exists a $\bar{n}\in \N$ such that for all $n \geq \bar{n}$ there exists a code $c\in\mc{C}(n)$ that satisfies:
\begin{eqnarray}
\PP_{\textsf{e}}(c)  = \PP_c\bigg(\Big|\Big|Q^n - \QQ \Big|\Big|_{\sf{v}}\geq \varepsilon\bigg) + \PP_c\bigg( U^n \neq \hat{U}^n \bigg) \leq \varepsilon.
\end{eqnarray}
\end{definition}
If the error probability $\PP_{\textsf{e}}(c)$ is small, the empirical frequency of symbols $(u,s,x,y,\hat{u}) \in \mc{U}\times \mc{S}\times\mc{X}\times \mc{Y} \times\mc{U}$ is close to the probability distribution $\QQ(u,s,x,y,\hat{u})$, \textit{i.e.} the sequence of symbols $(U^n,S^n,X^n,Y^n,\hat{U}^n)\in A_{\varepsilon}^{{\star}{n}}(\QQ)$ are jointly typical for target distribution $\QQ(u,s,x,y,\hat{u})$, with large probability. In that case,  sequences of symbols are coordinated empirically. 

\section{Channel State Correlated with the Source}\label{sec:CorrelGPsideInfoEnc}

We consider probability distributions $\PP_{\sf{us}}$ for the source  and $\mc{T}_{\sf{y|xs}}$ for the channel.  
We characterize the set of probability distribution $\QQ(u,s,x,y,\hat{u}) \in  \Delta(\mc{U} \times \mc{S}\times \mc{X} \times \mc{Y} \times \mc{U}  )$ that are achievable.

\begin{theorem}[Source Correlated with Channel State]$\qquad\qquad$\label{theo:CorrelGPsideInfoEnc}
$1)$ The joint probability distribution $\QQ(u,s,x,y,\hat{u})$ is achievable if and only if it decomposes as follows:
\begin{eqnarray}
\begin{cases}
&\QQ(u,s) = \PP_{\sf{us}}(u,s),\\
& \QQ(y|x,s) = \mc{T}(y | x ,s),\\
& \QQ(\hat{u}|u) = \UN(\hat{u}|u),\\
&Y -\!\!\!\!\minuso\!\!\!\!- ( X,S)    -\!\!\!\!\minuso\!\!\!\!-   U , \\
&\PP_{us}(u,s)   \otimes   \QQ(x|u,s) \otimes \mc{T}(y|x,s) \otimes \UN(\hat{u}|u) \text{ is achievable}.\nonumber
\end{cases}
\end{eqnarray}
$2)$ The probability distribution $\PP_{\sf{us}}(u,s)   \otimes   \QQ(x|u,s) \otimes \mc{T}(y|x,s) \otimes \UN(\hat{u}|u)$ is achievable if:
\begin{eqnarray}
\max_{\tilde{\QQ}\in \Q} \bigg( I(U , W ; Y  )  -  I(W ; S |U )   -  H(U  )  \bigg)  > 0 ,  \label{eq:CorrelGPsideInfoEnc1}
\end{eqnarray}
$3)$ The probability distribution $\PP_{\sf{us}}(u,s)   \otimes   \QQ(x|u,s) \otimes \mc{T}(y|x,s) \otimes \UN(\hat{u}|u)$ is not achievable if:
\begin{eqnarray}
\max_{\tilde{\QQ}\in \Q} \bigg( I(U , W ; Y  )  -  I(W ; S |U )   -  H(U  )  \bigg)  < 0 ,  \label{eq:CorrelGPsideInfoEnc2}
\end{eqnarray}
where $\Q$ is the set of distributions $\tilde{\QQ}   \in  \Delta(\mc{U}   \times \mc{S}  \times \mc{W} \times \mc{X} \times \mc{Y}  \times \mc{U}  )$ with auxiliary random variable $W$ that satisfies:
\begin{small}
\begin{eqnarray*}
\begin{cases}
\sum_{w\in \mc{W}} \tilde{\QQ}(u,s,x,w,y,\hat{u}) \\
\quad =  \PP_{\sf{us}}(u,s)   \otimes   \QQ(x|u,s) \otimes \mc{T}(y|x,s) \otimes \UN(\hat{u}|u) , \\
Y -\!\!\!\!\minuso\!\!\!\!- (X , S ) -\!\!\!\!\minuso\!\!\!\!-  (U , W ) .
\end{cases}
\end{eqnarray*}
\end{small}
Since $\PP_{\sf{us}}(u,s)$, $\QQ(x|u,s)$, $ \mc{T}(y|x,s) $,  $\UN(\hat{u}|u)$ are fixed, the set $\Q$ corresponds to the set of transition probability $\tilde{\QQ}_{\sf{w|usx}}$.
\end{theorem}

Sketchs of proof of Theorem \ref{theo:CorrelGPsideInfoEnc} are stated in Appendices \ref{sec:ProofAchievability} and \ref{sec:ProofConverse}. We refer to equation \eqref{eq:CorrelGPsideInfoEnc1} as "information constraint". 

Transition probability $\QQ(x|u,s)$ is the unique degree of freedom for the joint distribution $\QQ(u,s,x,y,\hat{u})$ since $\PP_{\sf{us}}(u,s)$, $ \mc{T}(y|x,s) $ and $\UN(\hat{u}|u)$ are the data of the problem. $\QQ(x|u,s)$ characterizes the empirical coordination of the channel inputs $X^n$ with sequences of source/state $(U^n ,S^n)$. This result applies to  optimization problems of objective functions $\nu : \mc{U} \times \mc{S} \times \mc{X}  \times \mc{Y}  \mapsto \R$ that depend simultaneously on symbols $(U,S,X,Y)$ of source and channel. 





\begin{example}[Minimal Distortion Source/Input Symbols]$\qquad\qquad\qquad\qquad\qquad\qquad$\label{theo:CorrelGPDistortion}
Consider distortion function $d: \mc{U} \times \mc{X} \mapsto \R$ for source/input symbols. The minimal distortion level $D^{\star}$ is given by:
\begin{eqnarray}
D^{\star} &=& \min_{\QQ_{\sf{x|us}} \in \mc{A}^{\star}} \E\bigg[ d(U,X)\bigg], \label{eq:CorrelGPDistortion1Z}
\end{eqnarray}
where $\mc{A}^{\star} $ is the convex closure of the set of achievable $\QQ_{\sf{x|us}}$:
\begin{tiny}
\begin{eqnarray}   
\mc{A}^{\star} &=& \texttt{Conv}\bigg\{ \QQ_{\sf{x|us}} \text{ s.t. }  \max_{\tilde{\QQ}_{\sf{w|usx}}} \Big( I(U , W ; Y)  -  I(W ; S |U )   -  H(U  )  \Big)  \geq 0\bigg\}.\label{eq:CorrelGPAchievT}
\end{eqnarray}
\end{tiny}
Equation \eqref{eq:CorrelGPDistortion1Z} is a direct consequence of Theorem \ref{theo:CorrelGPsideInfoEnc}, see also \cite{LeTreust(EmpiricalCoordination)14}. Consider a smaller distortion level $D^{\circ} < D^{\star}$. The corresponding transition probability $\QQ_{\sf{x|us}}^{\circ} \notin  \mc{A}^{\star}$ is not achievable. 
There is no  code $c(n)$ such that $X^n$ is coordinated, \textit{i.e.} jointly typical, with $(U^n,S^n)$ for target probability distribution $\PP_{\sf{us}} \otimes \QQ_{\sf{x|us}}^{\circ} $. Hence, distortion level $D^{\circ}$ is not achievable.
\end{example}
For any objective function $\nu$, Theorem \ref{theo:CorrelGPsideInfoEnc} characterizes the coordination $ \QQ_{\sf{x|us}} \in \mc{A}^{\star} $ that is achievable  and optimal \cite{LeTreust(EmpiricalCoordination)14}:
\begin{eqnarray}
\max_{\QQ_{\sf{x|us}} \in \mc{A}^{\star}} \E\big[ \nu(U,S,X,Y)\big].
\end{eqnarray}


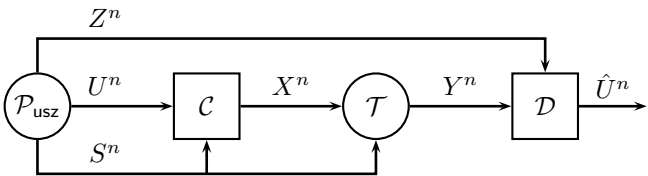
\begin{figure}[!ht]
\begin{center}
\psset{xunit=0.9cm,yunit=0.9cm}
\begin{pspicture}(0,-0.4)(8.5,1.8)
\pscircle(0,0.5){0.45}
\psframe(2,0)(3,1)
\pscircle(5,0.5){0.45}
\psframe(7,0)(8,1)
\psline[linewidth=1pt]{->}(0.5,0.5)(2,0.5)
\psline[linewidth=1pt]{->}(3,0.5)(4.5,0.5)
\psline[linewidth=1pt]{->}(5.5,0.5)(7,0.5)
\psline[linewidth=1pt]{->}(8,0.5)(9,0.5)
\psline[linewidth=1pt]{->}(0,0)(0,-0.5)(5,-0.5)(5,0)
\psline[linewidth=1pt]{->}(2.5,-0.5)(2.5,0)
\psline[linewidth=1pt]{->}(0,1)(0,1.5)(7.5,1.5)(7.5,1)
\rput[u](1,0.8){$U^n$}
\rput[u](1,-0.2){$S^n$}
\rput[u](1,1.8){$Z^n$}
\rput[u](3.75,0.8){$X^n$}
\rput[u](6.25,0.8){$Y^n$}
\rput[u](8.5,0.8){$\hat{U}^n$}
\rput(0,0.5){$\PP_{\sf{usz}}$}
\rput(2.5,0.5){$\C$}
\rput(5,0.5){$\mc{T}$}
\rput(7.5,0.5){$\D$}
\end{pspicture}
\caption{Decoder observes a side information $Z$ correlated with $(U,S)$.}
\label{fig:CorrelGPsideInfoEncZ}
\end{center}
\end{figure}
\vspace{-0.3cm}
Consider decoder observes a side information $Z$ correlated with $(U,S)$,  drawn i.i.d. with $\PP_{\sf{usz}}$, as stated in Fig. \ref{fig:CorrelGPsideInfoEncZ}. 

\begin{corollary}[Side Information at the Decoder]$\qquad\qquad$\label{theo:CorrelGPsideInfoEncZ}
Probability distribution $\PP_{\sf{usz}}(u,s,z)   \otimes   \QQ(x|u,s) \otimes \mc{T}(y|x,s) \otimes \UN(\hat{u}|u)$ is achievable if:
\begin{eqnarray}
\max_{\tilde{\QQ}\in \Q} \bigg( I(U , W ; Y,Z  )  -  I(W ; S |U )   -  H(U  )  \bigg)  > 0 ,  \label{eq:CorrelGPsideInfoEnc1Z}
\end{eqnarray}
where $\Q$ is the set of distributions $\tilde{\QQ}   \in  \Delta(\mc{U}   \times \mc{S}  \times \mc{Z}  \times \mc{W} \times \mc{X} \times \mc{Y}  \times \mc{U}  )$ with auxiliary random variable $W$ that satisfies:
\begin{small}
\begin{eqnarray*}
\begin{cases}
\sum_{w\in \mc{W}} \tilde{\QQ}(u,s,z,x,w,y,\hat{u}) \\
\quad =  \PP_{\sf{usz}}(u,s,z)   \otimes   \QQ(x|u,s) \otimes \mc{T}(y|x,s) \otimes \UN(\hat{u}|u) , \\
Y -\!\!\!\!\minuso\!\!\!\!- (X , S ) -\!\!\!\!\minuso\!\!\!\!-  (U , Z ,W ), \\
Z -\!\!\!\!\minuso\!\!\!\!- (U , S ) -\!\!\!\!\minuso\!\!\!\!-  (X , Y , W )  .
\end{cases}
\end{eqnarray*}
\end{small}
\end{corollary}
Only achievability result is stated in Corollary \ref{theo:CorrelGPsideInfoEncZ} but the converse holds. Proofs are obtained by replacing the channel output $Y$ by $(Y,Z)$ in App. \ref{sec:ProofAchievability} and \ref{sec:ProofConverse}.



\begin{corollary}[Removing Coordination of Channel Input]$\qquad\qquad$\label{theo:CorrelGPNoEC}
Lossless decoding is achievable if:
\begin{eqnarray}
\max_{\tilde{\QQ}_{\sf{xw|us}}} \bigg( I(U , W ; Y)  -  I(W ; S |U )   -  H(U  )  \bigg)  > 0 .  \label{eq:CorrelGPNoEC1Z}
\end{eqnarray}
\end{corollary}
Converse of Corollary \ref{theo:CorrelGPNoEC} also holds. As no coordination is required between $X$ and $(U,S)$, transition probability ${\QQ}_{\sf{x|us}}$ is a degree of freedom for maximizing \eqref{eq:CorrelGPNoEC1Z}. Proof is a direct consequence of Theorem \ref{theo:CorrelGPsideInfoEnc} since the maximum  in \eqref{eq:CorrelGPNoEC1Z} is taken over transitions ${\QQ}_{\sf{x|us}}$ and $\tilde{\QQ}_{\sf{w|usx}} $  instead of  $\tilde{\QQ}_{\sf{w|usx}} $ only.



 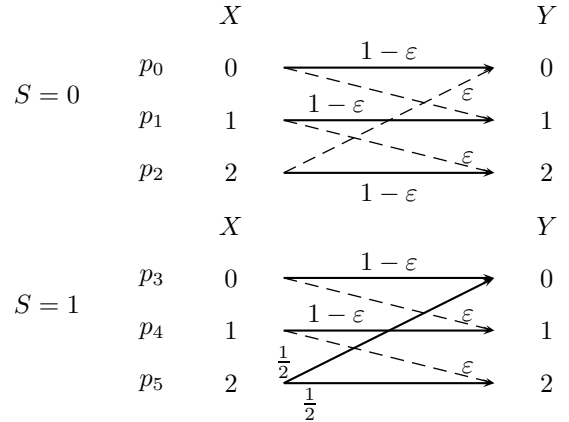
\begin{figure}[ht]
\begin{center}
\psset{xunit=0.7cm,yunit=0.7cm}
\begin{pspicture}(-1,-1)(15,2.5)
\rput(4,2.5){$X$}
\rput(4,1.5){$0$}
\rput(4,0.5){$1$}
\rput(4,-0.5){$2$}
\rput(2.5,1.5){$p_0$}
\rput(2.5,0.5){$p_1$}
\rput(2.5,-0.5){$p_2$}
\psline{->}(5,1.5)(9,1.5)
\psline{->}(5,0.5)(9,0.5)
\psline{->}(5,-0.5)(9,-0.5)
\psline[linewidth=0.5pt, linestyle = dashed]{->}(5,1.5)(9,0.5)
\psline[linewidth=0.5pt, linestyle = dashed]{->}(5,0.5)(9,-0.5)
\psline[linewidth=0.5pt, linestyle = dashed]{->}(5,-0.5)(9,1.5)
\rput(7,1.8){$1 - \varepsilon$}
\rput(7,-0.9){$1 - \varepsilon$}
\rput(6,0.8){$1 - \varepsilon$}
\rput(8.5,1){$\varepsilon$}
\rput(8.5,-0.2){$\varepsilon$}
\rput(10,2.5){$Y$}
\rput(10,1.5){$0$}
\rput(10,0.5){$1$}
\rput(10,-0.5){$2$}
\rput(0.5,1){$S = 0$}
\end{pspicture}
\begin{pspicture}(-1,-1)(15,3)
\rput(4,2.5){$X$}
\rput(4,1.5){$0$}
\rput(4,0.5){$1$}
\rput(4,-0.5){$2$}
\rput(2.5,1.5){$p_3$}
\rput(2.5,0.5){$p_4$}
\rput(2.5,-0.5){$p_5$}
\psline{->}(5,1.5)(9,1.5)
\psline{->}(5,0.5)(9,0.5)
\psline{->}(5,-0.5)(9,-0.5)
\psline{->}(5,-0.5)(9,1.5)
\psline[linewidth=0.5pt, linestyle = dashed]{->}(5,1.5)(9,0.5)
\psline[linewidth=0.5pt, linestyle = dashed]{->}(5,0.5)(9,-0.5)
\rput(7,1.8){$1 - \varepsilon$}
\rput(8.5,-0.2){$\varepsilon$}
\rput(6,0.8){$1 - \varepsilon$}
\rput(8.5,0.8){$ \varepsilon$}
\rput(5,-0.1){$\frac{1}{2}$}
\rput(5.5,-0.9){$\frac{1}{2}$}
\rput(10,2.5){$Y$}
\rput(10,1.5){$0$}
\rput(10,0.5){$1$}
\rput(10,-0.5){$2$}
\rput(0.5,1){$S = 1$}
\end{pspicture}
\caption{Channel $\mc{T}(y|x,s) $ with states $S\in\{0,1\}$ depending on error probability $\varepsilon \in [0,0.5]$. Probabilities $(p_0,p_1,p_2,p_3,p_4,p_5)$ characterize the transition probability $\QQ(x| s)$. The unique difference between states $S\in\{0,1\}$ holds for the channel input $X =2$. Exploitation of the correlation between channel state $S$ and source symbol $U$ improves the transmission.}\label{fig:CorrelBSAC2}
\end{center}
\end{figure}

\section{Comparison with Previous Work}\label{sec:PreviousWork}

In this section, we consider the result stated in Corollary \ref{theo:CorrelGPNoEC} for lossless decoding, removing the coordination between $X$ and $(U,S)$. 
We compare our result to the case of "State Amplification" \cite{KimSituvongCover08} and independent source/state $(U,S)$,  \cite{MerhavShamai03}.

 \begin{corollary}[Comparison with Previous Work]$\qquad\qquad\qquad\qquad\qquad$\label{coro:GelfandPinsker}
 $\bullet$ When the source is the state $S=U$, information constraint \eqref{eq:CorrelGPNoEC1Z} reduces to eq. (2) in \cite{KimSituvongCover08} for "state amplification":
\begin{eqnarray}
\max_{\tilde{\QQ}_{\sf{x|s}}} I(S , X ; Y  )     -  H(S ) > 0.
\end{eqnarray}
$\bullet$ When source $U$ is independent of state $S$, information constraint \eqref{eq:CorrelGPNoEC1Z} reduces to the one stated in  \cite{MerhavShamai03}, \cite{gelfand-it-1980}:
\begin{small}
\begin{eqnarray}
&\max_{\tilde{\QQ}_{\sf{xw|us}}} \bigg( I(U , W ; Y  )  -  I(W ; S |U )   -  H(U  )  \bigg)   > 0& \nonumber \\
&\Longleftrightarrow\max_{\tilde{\QQ}_{\sf{xw|s}}} \bigg( I( W ; Y  )  -  I(W ; S  )    -  H(U  ) \bigg)   >0.& \label{eq:GPconstraint}
\end{eqnarray}
\end{small}
\end{corollary}
Proof of equation \eqref{eq:GPconstraint} in Corollary \ref{coro:GelfandPinsker} is stated in \cite{LeTreust(RapportITW)14}. As mentioned in \cite{MerhavShamai03}, separation holds when random variables of the channel $(S,X,Y)$ are independent of the source $U$.

\begin{proposition}\label{prop:CorrelGap}
The benefit of exploiting the correlation between source $U$ and channel state $S$ is greater than:
\begin{footnotesize}
\begin{eqnarray}
&&\max_{\tilde{\QQ}_{\sf{xw|us}}} \bigg( I(U , W ; Y  )  -  I(W ; S |U )   -  H(U  )  \bigg)   \nonumber \\
&-& \max_{\tilde{\QQ}_{\sf{xw|s}}}  \bigg( I( W ; Y  )  -  I(W ; S  )   -  H(U  )  \bigg)  \geq  \min_{\tilde{\QQ}_{\sf{x|s}}} I(U  ; Y  ) .
\end{eqnarray}
\end{footnotesize}
\end{proposition}
Proof of Proposition \ref{prop:CorrelGap} is stated in \cite{LeTreust(RapportITW)14}.


\section{Numerical Results without Coordination}\label{sec:CorrelExampleBSAC}

We consider the channel with state $\mc{T}(y|x,s) $ depicted in figure \ref{fig:CorrelBSAC2} depending on  parameter $\varepsilon \in [0,0.5]$ and the joint probability distribution $\PP_{\sf{us}}(u,s)$ defined by figure \ref{fig:JointDistribution2}. 
 \begin{figure}[ht]
\begin{center}
\psset{xunit=0.7cm,yunit=0.7cm}
\begin{pspicture}(-1,0)(5,3.5)
\psframe(0,0)(4,3)
\psline(0,1.5)(4,1.5)
\psline(2,0)(2,3)
\rput(1,2.25){$\frac{1 + \alpha}{4}$}
\rput(3,2.25){$\frac{1 - \alpha}{4}$}
\rput(1,0.75){$\frac{1 - \alpha}{4}$}
\rput(3,0.75){$\frac{1 + \alpha}{4}$}
\rput(-0.85,2.25){$U=0$}
\rput(-0.85,0.75){$U=1$}
\rput(1,3.3){$S=0$}
\rput(3,3.3){$S=1$}
 \end{pspicture}
 \caption{Joint probability distribution $\PP_{\sf{us}}(u,s)$. If $\alpha = 0$, the random variables $U$ and $S$  are independent. If $\alpha = 1$, both random variables are equal: $U=S$. The marginal probability distributions on $U$ and $S$ are always equal to the uniform probability distribution $(0.5,0.5)$. } \label{fig:JointDistribution2}
\end{center}
\end{figure}
The correlation between the information source $U$ and the channel state $S$ depends on parameter $\alpha \in [0,1]$.
We evaluate the benefit of exploiting the correlation between source $U$ and channel state $S$ for different values of parameters  $\alpha\in [0,1]$ and $\varepsilon\in [0,0.5]$.
\begin{figure}[ht!]
\centering
\includegraphics[width=0.45\textwidth]{
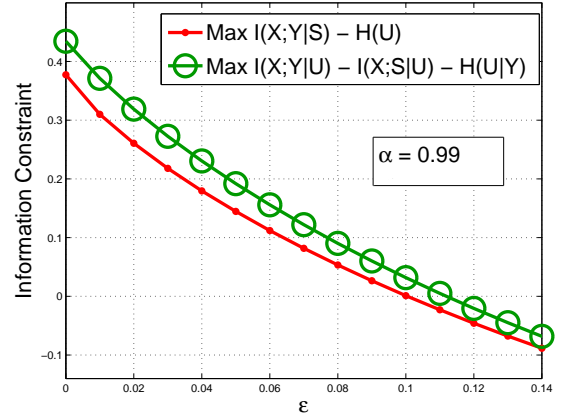}
\caption{Upper bound  \eqref{eq:UpperBoundGP} on \eqref{eq:GPconstraint}  is smaller than the lower bound \eqref{eq:LowerBoundCorrel} on information constraint \eqref{eq:CorrelGPNoEC1Z}  of Corollary \ref{theo:CorrelGPNoEC}.}
\label{fig:InformationConstraint_IXY_U_AE_figH_2014_04_29}
 \end{figure}
 Since the information constraint with auxiliary random variable $W$ is difficult to characterize, we consider the following upper \eqref{eq:UpperBoundGP} and lower \eqref{eq:LowerBoundCorrel} bounds:
\begin{footnotesize}
\begin{eqnarray}
  \text{\eqref{eq:GPconstraint}}  &\leq & \max_{(p_3,p_4,p_5) } \bigg( I( X ; Y  | S  )    -  H(U  ) \bigg),\label{eq:UpperBoundGP}\\
  \text{\eqref{eq:CorrelGPNoEC1Z}}  &\geq& \max_{(p_3,p_4,p_5) }  \bigg( I(X ; Y |U    )  -  I(X ; S |U )   -  H(U | Y  )  \bigg) . \label{eq:LowerBoundCorrel} 
\end{eqnarray}
\end{footnotesize}
\begin{itemize}
\item[$\bullet$] \eqref{eq:UpperBoundGP} is an upper bound on \eqref{eq:GPconstraint} that does not take into account the correlation between $U$ and $S$.
\item[$\bullet$] \eqref{eq:LowerBoundCorrel} is a lower bound on \eqref{eq:CorrelGPNoEC1Z} of Corollary \ref{theo:CorrelGPNoEC} that takes into account the correlation between $U$ and $S$.
\end{itemize} 
Only parameters $(p_3,p_4,p_5)$ are chosen in order to maximize the lower bound \eqref{eq:LowerBoundCorrel} on the information constraint. Regarding upper bound \eqref{eq:UpperBoundGP}, since parameters $(p_0,p_1,p_2) = (\frac{1}{3}, \frac{1}{3}, \frac{1}{3})$ maximize the mutual information $I( X ; Y  | S  =0) $, we only consider the maximization over parameters $(p_3,p_4,p_5) $. \\

$\bullet$ For $\alpha = 0.99$ in Fig. \ref{fig:InformationConstraint_IXY_U_AE_figH_2014_04_29}, the upper bound \eqref{eq:UpperBoundGP} on \eqref{eq:GPconstraint} is always smaller than the lower bound \eqref{eq:LowerBoundCorrel} on our information constraint \eqref{eq:CorrelGPNoEC1Z}. For some values of $\varepsilon$, the decoder can recover the source. Exploitation of the correlation between channel state $S$ and source symbol $U$ improves the transmission.  \\

\begin{figure}[ht!]
\centering
\includegraphics[width=0.45\textwidth]{
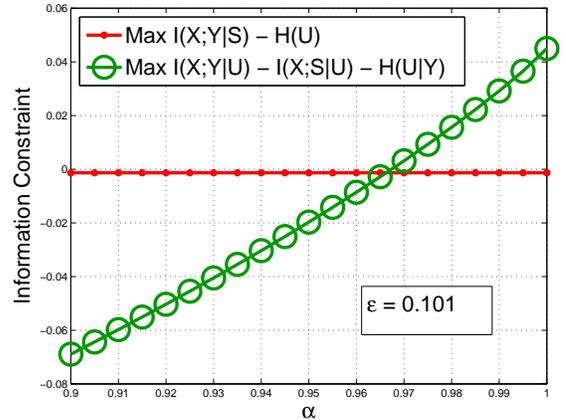}
\caption{Upper bound  \eqref{eq:UpperBoundGP} on \eqref{eq:GPconstraint} does not depend on $\alpha$ since the correlation between $U$ and $S$ is not taken into account.}
\label{fig:InformationConstraint_IXY_U_AE_figG_2014_04_29}
 \end{figure}

$\bullet$ For $\varepsilon = 0.101$ in Fig. \ref{fig:InformationConstraint_IXY_U_AE_figG_2014_04_29}, upper bound \eqref{eq:UpperBoundGP} on \eqref{eq:GPconstraint} is negative and does not depend on correlation parameter $\alpha$. 
For large values of $\alpha$, the \eqref{eq:LowerBoundCorrel} lower bound on \eqref{eq:CorrelGPNoEC1Z} is positive. Hence, the decoder can recover the information source exploiting the correlation between $U$ and $S$.


\section{Conclusion}\label{sec:Conclusion}

We consider a state dependent channel in which the channel state is correlated with the source symbol. We investigate simultaneously the lossless transmission of information and the empirical coordination of channel inputs with the symbols of source and states. We characterize the set of achievable empirical distributions with or without side information at the decoder. Based on this result, we determine the minimal distortion between symbols of source and channel inputs under lossless decoding constraint. Correlation between information source and channel state improves the feasibility of the lossless transmission.


\appendices

\section{Sketch of Achievability of Theorem \ref{theo:CorrelGPsideInfoEnc}}\label{sec:ProofAchievability}
There exists $\delta>0$, rates $\sf{R}_{\sf{M}} \geq 0$ and $\sf{R}_{\sf{L}} \geq 0$ such that :
\begin{eqnarray} 
\sf{R}_{\sf{M}}  &\geq&  H(U ) + \delta ,\label{eq:CorrelGPParameters1}\\
\sf{R}_{\sf{L}}  &\geq&  I(W ; S |U ) + \delta, \label{eq:CorrelGPParameters2} \\
\sf{R}_{\sf{M}} +  \sf{R}_{\sf{L}}   &\leq&  I(U , W ; Y ) - \delta  \label{eq:CorrelGPParameters3}.
\end{eqnarray}

\begin{itemize}
\item[$\bullet$] \textit{Random codebook.} Generate $|\mc{M} |= 2^{n    \sf{R}_{\sf{M}} } $ sequence $U^n(m)$ with index $m\in\mc{M}$ drawn from the i.i.d. marginale probability distribution $\PP_{\sf{u}}^{\otimes n}$. For each index $m \in \mc{M}$, generate $|\mc{M}_{\sf{L}}    |= 2^{n    \sf{R}_{\sf{L}} } $ sequences $W^n(m,l)\in \mc{W}^n $ with index $l \in  \mc{M}_{\sf{L}} $, drawn from the i.i.d. marginal probability distribution $\QQ_{\sf{w|u}}^{\otimes n}$ depending on $U^n(m)$.\\ 
\item[$\bullet$] \textit{Encoding function.}  The encoder observes the sequence of source symbols $U^n \in  \mc{U}^n$ and finds the index $m\in \mc{M}$ such that $U^n = u^n(m)$. It finds the index $l\in \mc{M}_{\sf{L}}$ such that the sequence  $w^n(m,l) \in A_{\varepsilon}^{{\star}{n}}(U^n,S^n)$ is jointly typical with the source sequences $(U^n,S^n)$. Encoder sends the sequence $X^n$ drawn from the probability $\QQ_{\sf{x|wus}}^{\otimes n}$ depending on sequences $(w^n(m,l),u^n(m), S^n)$.\\
\item[$\bullet$] \textit{Decoding function.} The decoder observes the sequence $Y^n$ and finds the indexes $(m,l) \in \mc{M}  \times \mc{M}_{\sf{L}} $ such that $( u^n(m) , w^n(m,l) ) \in A_{\varepsilon}^{{\star}{n}}(Y^n)$. It returns sequence $\hat{u}^n = u^n(m)$.\\
\end{itemize}

From Packing and Covering Lemmas stated in \cite{ElGammalKim(book)11} pp. 46 and 208, equations \eqref{eq:CorrelGPParameters1}, \eqref{eq:CorrelGPParameters2}, \eqref{eq:CorrelGPParameters3}
 imply the expected probability of error events are bounded by $\varepsilon$ for all $n\geq\bar{n}$:
\begin{tiny}
\begin{eqnarray}
&&\E_c\bigg[ \PP\bigg( \forall  m \in \mc{M} ,\quad U^n(m) \neq U^n \bigg)\bigg]  \leq \varepsilon,\\
&&\E_c\bigg[ \PP\bigg( \forall  l \in \mc{M}_{\sf{L}}  ,\quad W^n(m,l) \notin A_{\varepsilon}^{{\star}{n}}(U^n(m),S^n)\bigg)\bigg]  \leq \varepsilon,\\
&&\E_c\bigg[ \PP\bigg(  \exists (m',l')\neq  (m,l)  ,\text{ s.t. } (U^n(m'), W^n(m',l') ) \in A_{\varepsilon}^{{\star}{n}}(Y^n)    \bigg)\bigg] \leq \varepsilon,\\
&&\E_c\bigg[ \PP\bigg(  \exists m'\neq  m  ,\text{ s.t. } (U^n(m'), W^n(m',l) ) \in A_{\varepsilon}^{{\star}{n}}(Y^n)     \bigg)\bigg]   \leq \varepsilon.
\end{eqnarray}
\end{tiny}
There exists a code $c^{\star} \in \mc{C}(n)$ with $\PP_{\textsf{e}}(c^{\star})  \leq 8 \varepsilon$ for all $n\geq \bar{n}$. Lossless decoding condition is satisfied since $\PP_{c^{\star}}(U^n \neq \hat{U}^n) \leq 4 \varepsilon $. Sequences $(U^n,S^n,X^n,Y^n,\hat{U}^n) \in A_{\varepsilon}^{{\star}{n}}(\QQ)$ are jointly typical with high probability, hence empirical coordination requirement is satisfied.

\section{Sketch of Converse of Theorem \ref{theo:CorrelGPsideInfoEnc}}\label{sec:ProofConverse}
Define error event $E=1$ if $(U^n,S^n,X^n,Y^n,\hat{U}^n) \notin A_{\varepsilon}^{{\star}{n}}$ and $E=0$ otherwise.
\vspace{-0.3cm} 
\begin{tiny}
\begin{eqnarray}
&&n\cdot  H(U  ) \nonumber\\
&=& H(U^n  ) \label{eq:CorrelGPsideInfoEnc_1} \\
&\leq& H(U^n | E=0  ) + n \cdot   \varepsilon \label{eq:CorrelGPsideInfoEnc_2} \\
&=&  I(U^n ;  Y^n  |E = 0 )    + H(U^n |Y^n ,E=0 ) + n\cdot   \varepsilon   \label{eq:CorrelGPsideInfoEnc_11} \\
&\leq&  I(U^n ;  Y^n |E = 0 )    +  n\cdot   2\varepsilon  \label{eq:CorrelGPsideInfoEnc_3} \\
&=&  \sum_{i=1}^n I(U^n ;  Y_i |  Y^{i-1}, E=0 )    +  n\cdot   2\varepsilon  \label{eq:CorrelGPsideInfoEnc_4} \\
&\leq &  \sum_{i=1}^n I(U^n , Y^{i-1} ;  Y_i , E=0 )    +  n\cdot   2\varepsilon  \label{eq:CorrelGPsideInfoEnc_5} \\
&= &  \sum_{i=1}^n I(U^n , Y^{i-1} , S_{i+1}^n ;  Y_i | E=0 )  \nonumber\\
 &-&   \sum_{i=1}^n I( S_{i+1}^n ;  Y_i |  U^n , Y^{i-1} , E=0 )  +  n\cdot  2 \varepsilon  \label{eq:CorrelGPsideInfoEnc_6} \\
&= &  \sum_{i=1}^n I(U^n , Y^{i-1} , S_{i+1}^n ;  Y_i  |  E=0)   \nonumber\\
 &-&   \sum_{i=1}^n I( S_i ; Y^{i-1}  |  U^n ,  S_{i+1}^n , E=0 )  +  n\cdot   2\varepsilon  \label{eq:CorrelGPsideInfoEnc_7} \\
&= &  \sum_{i=1}^n I(U_i,U^{-i} , Y^{i-1} , S_{i+1}^n ;  Y_i |  E=0 )  \nonumber\\
 &-&  \sum_{i=1}^n I( S_i ;  U^{-i}, Y^{i-1}  , S_{i+1}^n | U_i,E=0 )  +  n\cdot   3\varepsilon
 \label{eq:CorrelGPsideInfoEnc_8} \\
&= &  \sum_{i=1}^n I(U_i, W_i  ;  Y_i  |  E=0)   -   \sum_{i=1}^n I( W_i  ;  S_i| U_i  ,E=0 )  +  n\cdot   3\varepsilon  \label{eq:CorrelGPsideInfoEnc_9} \\
&\leq &n \cdot  \max_{\tilde{\QQ}\in \Q}   \bigg( I ( U,W ; Y  )  -  I(W ; S | U)  \bigg)  +  n\cdot   4\varepsilon  \label{eq:CorrelGPsideInfoEnc_10} .
\end{eqnarray}
\end{tiny}
Eq. \eqref{eq:CorrelGPsideInfoEnc_1} comes from the i.i.d. property of the source.\\
Eq. \eqref{eq:CorrelGPsideInfoEnc_2} is due to Fano for coordination: $\PP(E=1)\leq \varepsilon $, \cite{LeTreust(RapportITW)14}.\\
Eq. \eqref{eq:CorrelGPsideInfoEnc_3} is due to Fano's inequality for lossless decoding \cite{LeTreust(RapportITW)14}.\\
Eq. \eqref{eq:CorrelGPsideInfoEnc_11}, \eqref{eq:CorrelGPsideInfoEnc_4}, \eqref{eq:CorrelGPsideInfoEnc_5}  and  \eqref{eq:CorrelGPsideInfoEnc_6} come from properties of MI.\\
Eq. \eqref{eq:CorrelGPsideInfoEnc_7} comes from the Csiszar sum identity.\\
Eq. \eqref{eq:CorrelGPsideInfoEnc_8} is due to  $(U_i,S_i)$ independent of $(U^{-i} ,  S_{i+1}^n )$,  \cite{LeTreust(RapportITW)14}.\\
Eq. \eqref{eq:CorrelGPsideInfoEnc_9} is due to auxiliary RV, $W_i = (U^{-i}, Y^{i-1}  , S_{i+1}^n)$ .\\
Eq. \eqref{eq:CorrelGPsideInfoEnc_10} is due to continuity of MI and maximum over auxiliary random variables $W$ s.t.  $Y -\!\!\!\!\minuso\!\!\!\!- (X , S ) -\!\!\!\!\minuso\!\!\!\!-  W$,  \cite{LeTreust(RapportITW)14}.


\begin{thebibliography}{10}

\bibitem{gelfand-it-1980}
S.~I. Gel'fand and M.~S. Pinsker, ``Coding for channel with random
  parameters,'' {\em Problems of Control and Information Theory}, vol.~9, no.~1,
  pp.~19--31, 1980.

\bibitem{costa-it-1983}
M.~H.~M. Costa, ``Writing on dirty paper,'' {\em IEEE Transactions on
  Information Theory}, vol.~29, no.~3, pp.~439--441, 1983.

\bibitem{heegard-it-1983}
C.~D. Heegard and A.~A. {El Gamal}, ``On the capacity of computer memory with
  defects,'' {\em IEEE Transactions on  Information Theory}, vol.~29, no.~5, pp.~731--739,
  1983.

\bibitem{MoulinOSullivan03}
P.~Moulin and J.~O'Sullivan, ``Information-theoretic analysis of information
  hiding,'' {\em IEEE Transactions on  Information Theory}, vol.~49, no.~3,
  pp.~563--593, 2003.
  
  \bibitem{LeTreust(EmpiricalCoordination)14}
M.~Le~Treust, ``Empirical coordination for the joint source-channel coding
  problem,'' {\em submitted to IEEE Trans. on Information Theory}, 2014.


\bibitem{KimSituvongCover08}
Y.-H. Kim, A.~Sutivong, and T.~Cover, ``State amplification,'' {\em IEEE Trans. on  Information Theory}, vol.~54, no.~5, pp.~1850--1859, May 2008.

\bibitem{WeissmanOrdentlich05}
T.~Weissman and E.~Ordentlich, ``The empirical distribution of rate-constrained
  source codes,'' {\em IEEE Transactions on  Information Theory}, vol.~51,
  no.~11, pp.~3718 -- 3733, 2005.

\bibitem{KramerSavari07}
G.~Kramer and S.~Savari, ``Communicating probability distributions,'' {\em
IEEE Trans. on  Information Theory}, vol.~53, no.~2, pp.~518 -- 525,
  2007.

\bibitem{CuffPermuterCover10}
P.~Cuff, H.~Permuter, and T.~Cover, ``Coordination capacity,'' {\em IEEE Trans. on  Information Theory}, vol.~56, no.~9, pp.~4181--4206, 2010.


\bibitem{ElGammalKim(book)11}
A.~E. Gamal and Y.-H. Kim, {\em Network Information Theory}.
\newblock Cambridge University Press, Dec. 2011.

\bibitem{MerhavShamai03}
N.~Merhav and S.~Shamai, ``On joint source-channel coding for the wyner-ziv
  source and the gel'fand-pinsker channel,'' {\em IEEE Transactions on
  Information Theory}, vol.~49, no.~11, pp.~2844--2855, 2003.

\bibitem{LeTreust(RapportITW)14}
M.~Le~Treust, ``Correlation between channel state and information source with
  empirical coordination constraint,'' tech. rep.,
  https://sites.google.com/site/maelletreust/B.pdf?attredirects=0\&d=1, 2014.

\end{thebibliography}
\end{document}